\documentclass[journal, twocolumns]{IEEEtran}
\usepackage[utf8]{inputenc}
\usepackage{soul}
\usepackage{color}
\usepackage{graphics}
\usepackage{bm}
\usepackage{graphicx}
\usepackage{mathtools}
\usepackage{amsmath}
\usepackage[linesnumbered,ruled]{algorithm2e}
\usepackage{algpseudocode}
\usepackage[acronym,shortcuts]{glossaries}
\usepackage{subfigure}
\usepackage{balance}
\usepackage{siunitx}
\usepackage{setspace}
\usepackage{tikz}
\usepackage{textcomp}
\usepackage{hyperref}
\usepackage{lipsum}
\usepackage{multirow}

\newacronym{ris}{RIS}{Reflective Intelligent Surface}
\newacronym{uav}{UAV}{Unmanned Aerial Vehicle}
\newacronym{fpga}{FPGA}{Field-Programmable Gate Array}
\newacronym{pin}{PIN}{Positive-Intrinsic-Negative}
\newacronym{mems}{MEMS}{Micro Electro Mechanical System}
\newacronym{fets}{FETs}{Field-Effect Transistors}
\newacronym{dc}{DC}{Direct Current}
\newacronym{rf}{RF}{Radio Frequency}
\newacronym{udn}{UDN}{Ultra-Dense Network}
\newacronym{iot}{IoT}{Internet of Things}
\newacronym{mimo}{MIMO}{Multiple-Input Multiple-Output}
\newacronym{mmwave}{mmWave}{Millimeter Wave}
\newacronym{5g}{B5G}{Beyond Fifth-Generation}
\newacronym{bs}{BS}{Base Station}
\newacronym{ser}{SER}{Symbol Error Rate}
\newacronym{los}{LoS}{Line-of-Sight}
\newacronym{lap}{LAP}{Low-Altitude Platform}
\newacronym{hap}{HAP}{High-Altitude Platform}
\newacronym{rbf}{RBF}{Radial Base Function}
\newacronym{ula}{ULA}{Uniform Linear Array}
\newacronym{snr}{SNR}{Signal-to-Noise-Ratio}
\newacronym{sca}{SCA}{Successive Convex Approximation}
\newacronym{pso}{PSO}{Particle Swarm Optimization}
\newacronym{pls}{PLS}{Physical Layer Security}

\title{Unmanned Aerial Vehicles  \\ Meet Reflective Intelligent Surfaces \\ to Improve Coverage and Secrecy}
\author{A. Brighente, M. Conti, H. Idriss, and S. Tomasin

\thanks{A. Brighente and M. Conti are with Department of Mathematics and HIT Research Center, University of Padova, Italy.}%
\thanks{H. Idriss is with the Lebanese International University, Lebanon.}%
\thanks{S. Tomasin is with the Department of Information Engineering, University of Padova, Italy.}
}
\date{}

\usepackage[normalem]{ulem}
\usepackage[user]{zref}
\newcommand{\sot}[1]{}

\newcounter{revc}
\makeatletter \zref@newprop{revcontent}{} \zref@addprop{main}{revcontent}
\zref@newprop{revsec}{} \zref@addprop{main}{revsec}
\zref@newprop{revpage}{} \zref@addprop{main}{revpage}

\newcommand{\revi}[2]{
\zref@setcurrent{revsec}{\thesection}%
\zref@setcurrent{revpage}{\thepage}%
\zref@setcurrent{revcontent}{#2}%
\refstepcounter{revc}%
\label{#1}%
\zlabel{#1}%
\uline{#2}%
}

\newcommand{\revinu}[2]{%
\zref@setcurrent{revsec}{\thesection}%
\zref@setcurrent{revcontent}{#2}%
\refstepcounter{revc}%
\zlabel{#1}%
\label{#1}
#2 }

\newcommand{\revr}[2]{%
\zref@setcurrent{revsec}{\thesection}%
\zref@setcurrent{revcontent}{#2}%
\refstepcounter{revc}%
\zlabel{#1}%
\label{#1} \sot{#2}} \makeatother

\newcommand\copyrighttext{%
  \footnotesize \textcopyright 2012 IEEE. Personal use of this material is permitted.
  Permission from IEEE must be obtained for all other uses, in any current or future
  media, including reprinting/republishing this material for advertising or promotional
  purposes, creating new collective works, for resale or redistribution to servers or
  lists, or reuse of any copyrighted component of this work in other works.}
\newcommand\copyrightnotice{%
\begin{tikzpicture}[remember picture,overlay]
\node[anchor=south,yshift=10pt] at (current page.south) {\fbox{\parbox{\dimexpr\textwidth-\fboxsep-\fboxrule\relax}{\copyrighttext}}};
\end{tikzpicture}%
}

\begin{document}

\maketitle
\copyrightnotice
\begin{abstract}
The high configurability and low cost of \acp{ris} made them a promising solution for enhancing the capabilities of \ac{5g} networks. Recent works proposed to mount \acp{ris} on \acp{uav}, combining the high network configurability provided by \ac{ris} with the mobility brought by \acp{uav}. However, the \ac{ris} represents an additional weight that impacts the battery lifetime of the \ac{uav}. Furthermore, the practicality of the resulting link in terms of communication channel quality and security have not been assessed in detail. In this paper, we highlight all the essential features that need to be considered for the practical deployment of \ac{ris}-enabled \acp{uav}. We are the first to show how the \ac{ris} size and its power consumption impact the \ac{uav} flight time. We then assess how the \ac{ris} size, carrier frequency, and \ac{uav} flying altitude affects the path loss. Lastly, we propose  a novel particle swarm-based approach to maximize coverage and improve the confidentiality of transmissions in a cellular scenario with the support of \acp{ris} carried by \acp{uav}. 
\end{abstract}
\glsresetall
\section{Introduction}
Massive \ac{mimo} represents the key technology employed by \ac{5g} wireless networks to serve billions of active wireless devices which are still increasing annually. In conjunction with \acp{mmwave}, massive \ac{mimo} enables \ac{5g} networks to support novel uses cases, such as virtual and augmented reality, high-fidelity holographic projections, digital twins, and autonomous driving. All these scenarios can meet the large bandwidth available at higher frequencies. Although literature widely discussed applications and advantages of the \ac{mmwave} technology, its widespread adoption is still limited by both its severe path-loss and its sensitivity to blockages. Despite of the good achievable rate, \ac{mimo} and \ac{mmwave} require complex signal processing algorithms, are costly, and present high energy demand to support an high number of Radio Frequency chains. \ac{udn} is another paradigm that has been proposed for \ac{5g} systems. \ac{udn} technology envisions the deployment of a large number of \acp{bs} that are not only expensive in terms of hardware and maintenance, but also increase the amount of interference in the network. \acp{ris} represent a new energy-efficient technology for re-configuring the wireless propagation environment via software-controlled reflections. This technology falls under different names in the literature, such as Reconfigurable Intelligent Surface, or Intelligent Reflective Surfaces, which however represent the same type of devices. \acp{ris} improve the performance of wireless communication networks by using massive low-cost passive reflecting elements integrated on a surface, therefore eliminating the need for \ac{rf} chains. By independently controlling the amplitude and phase of each element, \acp{ris} exploit passive beamforming to either enhance the signal directivity or null signals at the receivers. Furthermore, \acp{ris} inherently enable full-duplex communications without interference, providing higher spectral efficiency than other solutions. 

To fully exploit the potential of \acp{ris}, recent works focused on enhancing the flexibility of the network by moving \acp{ris} in different locations over time. Therefore, the \ac{ris} shall be mounted on movable devices with a certain automation level and able to coordinate with other devices in the network. A relevant example of such devices is given by \acp{uav}. When moving communication equipment's operating as \acp{bs}, relays, or \acp{ris}, \acp{uav} enhance the performance of wireless networks due to their high mobility, low cost, low power consumption, and high mobility. This framework can be exploited for several purposes, as we later discuss throughout the paper. However, using \ac{ris}-enabled \ac{uav} opens new challenges compared to regular \acp{uav}. In fact, the \ac{ris} controller demands for power from the \ac{uav} battery, shortening its lifetime. Furthermore, the weight of the \ac{ris} requires extra thrust from the \ac{uav} to fly, further shortening the battery lifetime.

In this paper, we analyze the requirements and limits of the deployment of \ac{ris}-enabled \acp{uav}. We first provide a characterization of \acp{ris} in terms of weight and power demand, to show how they impact on the flight performance of a \ac{uav}. Furthermore, since the flying altitude of \acp{uav} depends on the countries' policy, we show how the altitude impacts communications in terms of path loss incurred in the communication channel transmitter-\ac{ris}-receiver. Thanks to this characterization, we then show the performance of \ac{ris}-enabled \acp{uav} for two different applications in cellular networks: i) maximization of the average users' achievable rate, and ii) minimization of the network area secrecy to prevent eavesdropping by malicious users. In particular, we provide the following contributions.
\begin{itemize}
    \item We investigate the fundamental limits of flight time of UAVs carrying RIS. First, we derive the impact of the RISs  on the UAV power consumption, which in turn significantly reduces the flight time. Hence, we conclude that it is fundamental to design proper energy harvesting schemes or use backup drones to guarantee the network's resiliency.
    \item We investigate the capabilities of a RIS-enabled UAV in the context of millimeter wave communications. To optimize the spectral efficiency, we design a novel particle swarm optimization algorithm for the optimal deployment of drones and a coordinate descent algorithm to compute the optimal power allocation and RIS phase-shift configuration. By accounting for the RIS path loss modeling at those frequencies, we show how UAV's flight altitudes imposed by real regulations in different countries and the number of RIS elements impact the network performance.
    \item By using the optimization strategies of the previous point, we optimize the secrecy capacity of the network in a predefined area. We investigate the impact of the UAV's altitude and the RIS size on the network secrecy performance.
\end{itemize} 
We believe our paper can serve as a reference for the design of networks exploiting  \ac{ris} mounted on \ac{uav}, providing insights on both technological limitations and benefits.

\section{Reflective Intelligent Surfaces}
\label{two}
In this section, we first discuss the \ac{ris}'s architecture and dimension, and then provide an overview of their power consumption. 

The typical hardware architecture of a \ac{ris} is based on the concept of meta-surface, which is a planar array consisting of a large number of meta-atoms. A \ac{ris} consists of three layers and a controller. The first layer (outer layer) consists of a metallic patch of reflecting elements printed on a dielectric substrate to directly interact with the incident signals. The second layer (middle layer) is a copper plate used to avoid leakage in the signal energy. Lastly, the third layer (inner layer) is a control circuit board responsible for adjusting the reflection amplitude and phase shift of each element; this layer is triggered by a smart controller attached to the \ac{ris}. Due to their passive nature, \acp{ris} can be fabricated with limited layer thickness. Therefore, their weight is limited making them suitable to be mounted on different surfaces. A \ac{fpga} can operate as controller, which also helps in the communication and coordination with other network components. Phase and/or amplitude adjustment can be obtained using \ac{pin} diodes, \ac{fets}, or \ac{mems} switches \cite{r5}. 
Each element includes a \ac{pin} diode, which can be switched ON and OFF by controlling its biasing voltage using the controller, thereby generating a phase shift. A resistor with variable load is applied to control the reflection amplitude. By changing the value of this resistor, different portions of the incident signal’s energy are dissipated, thus the reflection amplitude varies between [0, 1]. The number of elements in a \ac{ris} relatively affects amplification of signals. The size $\lambda$ of the elements is in the order of fractions of the wavelength of the transmission carrier. Common values vary between  $\lambda/10$ and  $\lambda/2$ with a spacing of $\lambda/2$. For instance, the dimension of a $20\times 30$ \ac{ris} with a $28$~GHz ($\lambda = 1$~cm) carrier frequency is $0.2 \times 0.3$~m.
The weight of the \ac{ris} depends on the three layers. Based on the previous discussion, the heavier layer is the copper plate, whereas the controller and reflective elements do not significantly contribute. Very few works discussed physical implementations of \acp{ris} \cite{r5, r14, r15}. We show the details of these implementations in Table \ref{tab:risSpec}.

As previously mentioned, one of the benefits of \acp{ris} is their low energy consumption. The total power consumption of the \ac{ris} at a certain time instant depends on the number of active elements (i.e., the number of diodes in the ON state) and the power needed by the specific employed controller. Thus, from the elements point of view, the \ac{ris} power consumption is given by the summation of the power consumed by each active \ac{pin} diode. 

\begin{table}[!t]
    \centering
     \caption{Specifications of Different Implemented \acp{ris}}
    \begin{tabular}{|p{0.5in}|p{0.8in}|p{0.4in}|p{0.6in}|p{0.2in}|p{0.4in}|p{0.1in}|}
    \hline
 Operating Frequency [GHz] & Number of Reflecting Elements & \ac{ris} Area [m$^2$] & Power Consumption [W] & Ref. \\
 \hline
        2.4  & 6$\times$8 = 48 &  0.015 & not available & \cite{r14}  \\ \hline
        4.25  & 8$\times$32 = 256 & 0.037  & 1.28  & \cite{r5}  \\ \hline
        5 & 8$\times$8 = 64 &  0.004  & not available & \cite{r15}  \\ \hline
        10.5  & 100$\times$102 = 10200  &  1.02  & 10.56  & \cite{r5}  \\ \hline
        10.5  & 50$\times$34 = 1700 &  0.17  & 10.56   & \cite{r5}  \\ \hline    
    \end{tabular}
    \label{tab:risSpec}
\end{table}

\section{RIS-Enabled UAVs}
\label{three}
In this section, we provide an overview of the applications of \ac{ris}-enabled \acp{uav}. We first discuss the physical limits  UAVs that may impact on the network performance. Then we describe two scenarios where \ac{ris}-enabled \acp{uav} may provide significant advantages.

\subsection{A Review of the UAV Technology}
\acp{uav} are a technology in vogue and are envisioned to be employed for several purposes, such as aerial photography for journalism and movie, express shipping and delivery, and wireless communications. In this paper, we focus on the use of \acp{uav} to create air-ground \ac{los} links and enhance communication network performance. In this case, a \ac{ris}-enabled \ac{uav} is goes over a route to serve the network users. The optimization problem of finding the most suitable \ac{uav}'s location at any time is solved by a ground controller. The resulting quality of the communication links highly depends on the features of the employed \ac{uav}. In fact, according to the physical features of the \ac{uav}, the controller either creates channels used for a long time period (e.g., when hovering on a certain place), or open up opportunistic windows for one-time transmissions.

In this context, an key performance indicator is the flight time, defined as the maximum time a \ac{uav} can spend flying without getting recharged or refueled. This time ranges from tens of minutes to few hours, depending on the model and make and is also influenced by the weight and size of the load. In fact, the payload adds an additional weight to the \ac{uav} structure, and the \ac{uav} engine has to generate additional thrust to compensate for it. Furthermore, the size, shape, and position of the load influence the \ac{uav} flight time. For example, a wide load acts as a sail in the presence of wind, requiring the stability control module of the \ac{uav}  to spend more energy in compensating for the additional force imposed by the sail: this reduces the flight time.

\subsection{Applications of RIS-enabled UAVs}
\ac{ris}-enabled \acp{uav} provide a highly configurable solutions for extending the basic two-dimensional network model to the third dimension, fulfilling one of the requisites of future generation networks. The application of this technology spans different domains~\mbox{\cite{abdalla20}}.

\textbf{Creation of Communication Links}. When considering \ac{mmwave} communications, the presence of a \ac{los} link  significantly improves the link performance. On the contrary,  blockage may bring to service unavailability due to the high attenuation at \ac{mmwave} frequencies that strongly limits the propagation by reflection and scattering. When no \ac{los} channel is present, the \ac{ris}-enabled \ac{uav} may be deployed to create a communication link~\cite{r1, r3}. In this case, the \ac{ris} reflects the incident signal to the desired location, whereas the \ac{uav} provides the flexibility needed in a dynamic environment.

\textbf{Increased Coverage and Capacity}. A \ac{ris} enabled \ac{uav} can be optimally located to assist cell-edge users that may not be directly reachable by the \ac{bs}~\cite{r2}. Indeed, the signal transmitted by the \ac{bs} impinges the \ac{ris} elements which, via beamforming, irradiate a signal reaching cell-edge users. Another solution to improve coverage provides the adoption of Intelligent Omni-Surfaces, i.e., particular \acp{ris} having antenna elements on both sides of the meta-surface, thus capable of reflecting signals on both sides: when mounted vertically, they enable reflections in two areas of the cell. Irrespectively of the employed \ac{ris} technology and thanks to the 3D mobility of \acp{uav}, it is possible to extend the cell coverage and to increase the network capacity in a dynamic set-up. In the latter case, \ac{ris}-enabled \ac{uav} can overcome the limitations imposed by blockages by dynamically adjusting the location of the \ac{ris} or by moving it to areas with a high user concentration to help in the resource allocation process.

\textbf{Wireless Information and Power Transfer}
The large number of \ac{iot} devices in future networks is  challenging for energy management. Indeed, all these devices are battery-powered and there is a fundamental need for smart solutions to reduce their power consumption or provide external power supplies. To this end, wireless power transfer enables the charging of batteries without wired connections and \ac{ris}-assisted \ac{uav} may  support  the charging process. In fact, thanks to the high mobility provided by \acp{uav} and the low-energy demand of \acp{ris} it is possible to automate  charging  and convey power to multiple \ac{iot} nodes while at the same time also providing energy to   the \ac{uav}. During the charging process, it is also possible to encode information in the signal that delivers the power in what is commonly known as wireless information and power transfer. Thus, \ac{ris}-enabled \acp{uav} not only  dynamically deliver power to \ac{iot} nodes, but at the same time collect the information they generate, hence optimizing the network resources.
    
\textbf{Enhanced \ac{pls}}. A \ac{ris}-enabled \ac{uav} provides a configurable network that can be  prevent attacks such as eavesdropping thanks to directional jamming signals. The \ac{uav} mobility enables the optimal deployment of the \ac{ris} to optimize the  secrecy rate. Furthermore, when  the attacker can be identified, the \ac{uav} can  track it to provide continuous secrecy to the communications. Lastly, by using Intelligent Omni-Surfaces it is possible to control the secrecy rate over a pre-defined area. The optimal deployment of \ac{ris}-enabled \acp{uav} can also be exploited for physical layer authentication, providing dynamic access to users to the network's services when needed.

\section{Applicability of RIS-enabled UAVs}
Before deploying \ac{ris}-enabled \acp{uav} to serve a communication network, key features such as the path loss of the resulting channel and the maximum flight time needs to be considered. In this section we provide a numerical evaluation of the path loss incurred by the communication channel ground-\ac{ris}-ground, considering different carrier frequencies and \ac{ris}'s sizes. We provide the results considering the flying policies of different countries, which regulate the \ac{uav} flying altitude. Furthermore, we will show how the \ac{ris} size impacts the flight time of different \ac{uav} models. Lastly, we provide a case study on the effectiveness of \ac{ris}-enabled \acp{uav} in supporting security to the network. Here instead we consider the impact of the \ac{ris} size on both the security and the flight time of the \ac{uav}.

\subsection{Path Loss Modeling}
We consider a \ac{ris} of negligible thickness on an horizontal plane. Furthermore, we consider transmitter and receiver are in the far field with respect to the \ac{ris}.

\begin{figure}
    \centering
    \includegraphics[width = \columnwidth]{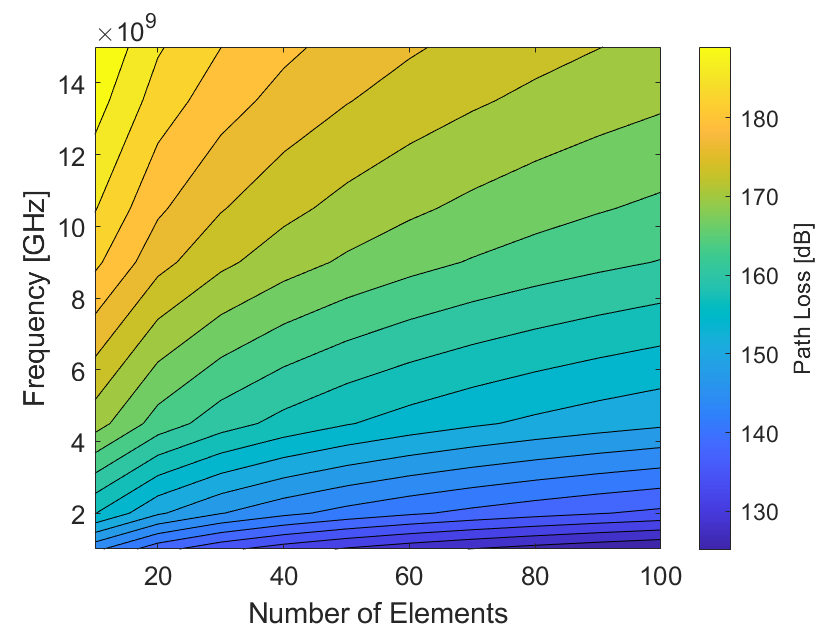}
    \caption{Path loss of the user-\ac{ris}-\ac{bs} channel for a \ac{uav} height of $50$~m. }
    \label{fig:pathloss_all}
\end{figure}

We compute the path loss according to the model in \cite{plModel}, focusing on the transmission frequency and the area of the \ac{ris}. Fig. \ref{fig:pathloss_all} shows the variation of path loss with respect to the transmission frequency and the number of elements of a \ac{ris}. For an \ac{uav}'s flying altitude of $50$~m. We notice that, irrespectively of the transmission frequency, increasing the number of elements reduces the path loss. On the other hand, for a given number of elements, the path loss increases with the carrier frequency. Therefore, to be able to exploit the advantages brought by \ac{mmwave} and THz frequencies, \acp{ris} shall be equipped with a large number of elements to avoid incurring in a prohibitively high path loss. However we notice that a higher number of elements also means a higher \ac{ris} size and therefore a higher weight to be carried by the \ac{uav}. Although usually the size of the elements depends on the carrier frequency, the inverse proportionality between these two measures is not sufficient to cope with the increase in path loss. Therefore, although elements are smaller for higher frequencies, the required number of elements still renders \acp{ris} larger at higher frequencies.


\subsection{\ac{ris}-Enabled \ac{uav} Flight Time}
\begin{figure}[!t]
    \centering
    \includegraphics[width=\columnwidth]{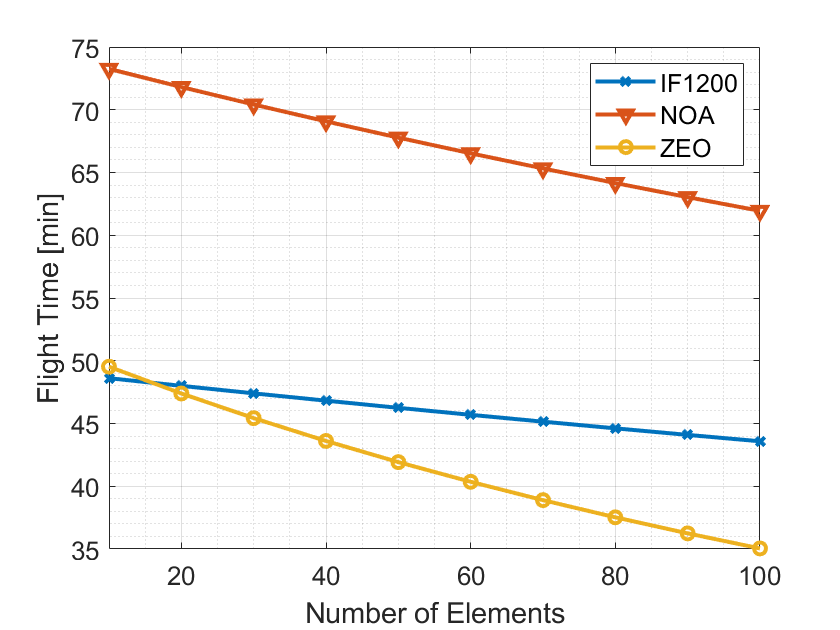}
    \caption{Variation of the flight time of different \acp{uav} with respect to the number of \ac{ris} elements.}
    \label{fig: flighttime}
\end{figure}

As we previously stated, the \ac{uav} flight time depends, among other factors, on the weight of the payload. Furthermore, when considering a \ac{ris}-enabled \ac{uav}, additional power is required to control the \ac{ris}. The additional power is related to both the activation of elements and powering the controller. 
We here show how this affetcs the flight time of different \ac{uav} models. Fig. \ref{fig: flighttime} shows the flight time vs. the number of \ac{ris} elements for \acp{uav} ZEO X4 \cite{r8}, Noa 6 \cite{r11}, and IF1200 \cite{r10}. Our choice on the \ac{uav} models is motivated by the information made available by the producers. We model the flight time and power consumption according to \cite{r123}.  With the increased number of elements the size of the \ac{ris} increases, and consequently its weight. We here consider a carrier frequency of $10$ GHz, and an \ac{ris} area going from $9 \cdot 10^{-3}$~m$^2$ to $9 \cdot 10^{-2}$~m$^2$. Due to the additional weight, the flight time decreases when increasing the area of the \ac{ris}. The most severe effect is obtained considering the ZEO, where the flight time decreases from $50$ min to $35$ min. Therefore, although an increased \ac{ris} size leads to smaller path loss values, the flight time significantly decreases. This highly impacts scenarios where the \ac{uav} needs to deliver a service hovering for a long time, or the \ac{uav} needs to fly a long path to reach its destination.

\subsection{Applications to \ac{pls}}
Due to their high configurability, \acp{ris} represent a fruitful technology for \ac{pls}. For instance, by suitably placing a \ac{ris}, we can improve the communication link to the legitimate receiver and at the same time alternate or null the link to an eavesdropper, assuming to know its channel. In a scenario where multiple users shall be securely served, multiple \acp{ris} would be particularly useful, but with high costs. The use of a \ac{ris}-enabled \ac{uav} reduces the deployment costs and provides a flexible solution, thank to the \ac{uav} freedom of movement. Therefore, although a \ac{ris}-enabled \ac{uav} may require a higher initial investment, it gives economic benefits in the long term, as it does not require  network reconfiguration when moving or deploying new \ac{ris}.

A researcher trend investigates  the advantages for \ac{pls} provided by \ac{ris}-enabled \acp{uav}. Authors in~\mbox{\cite{r1}} considered the uplink of several single-antenna users that communicate with a BS overa no \ac{los} channel: they therefore propose to exploit an \ac{ris}-enabled \ac{uav} to convey a secret signals to the BS  by jointly optimizing the trajectory of the UAV, the phase shifts of the RIS, and the user’s transmit power. Considering a multi-antenna \ac{bs} communicating to a single-antenna user, the authors in~\mbox{\cite{r2}} exploit a RIS-enabled UAV for the downlink: by optimizing the phase shifts of the RIS and the beamforming vector of the \ac{bs}, the system’s total energy efficiency is improved.  Authors in~\mbox{\cite{r3}} consider a non-LOS channel and exploit a \ac{ris}-enabled \ac{uav} to create a communication link:  the \ac{uav}, in addition to holding the \ac{ris}, plays the role of a relay between the transmitter and the receiver to create a self-interference at the \ac{uav}. By optimizing the number of reflecting elements and the altitude of the \ac{uav}, good performance in terms of outage probability, ergodic capacity, and energy efficiency are achieved.  The authors in~\mbox{\cite{r4}} considered a downlink communication, where the \ac{ris}-enabled \ac{uav} reflects the received signals from the \ac{bs} to other \acp{uav} at lower altitudes, which in turn broadcast the signal to ground users in their assigned area:  the performance of the symbol error rate and outage probability of the \ac{ris}-assisted UAV-UAV communications is studied.  

\section{Applications Performance}
\label{four}
In this section, we consider two examples of applications of \ac{ris}-enabled \acp{uav} to communication networks. We design a novel algorithm to optimally control the \ac{ris} phase shifts and the \ac{uav} location to maximize either the spectral efficiency (first case study) or the average secrecy rate (second case study). In the first application, we use the \acp{ris} to improve the coverage, here measured by the total coverage capacity. The second example pertains \ac{pls} as we impose confidential transmissions with respect to an eavesdropper in an unknown location. 

\subsection{Increased Coverage}
\begin{figure}[!t]
    \centering
    \includegraphics[width=\columnwidth]{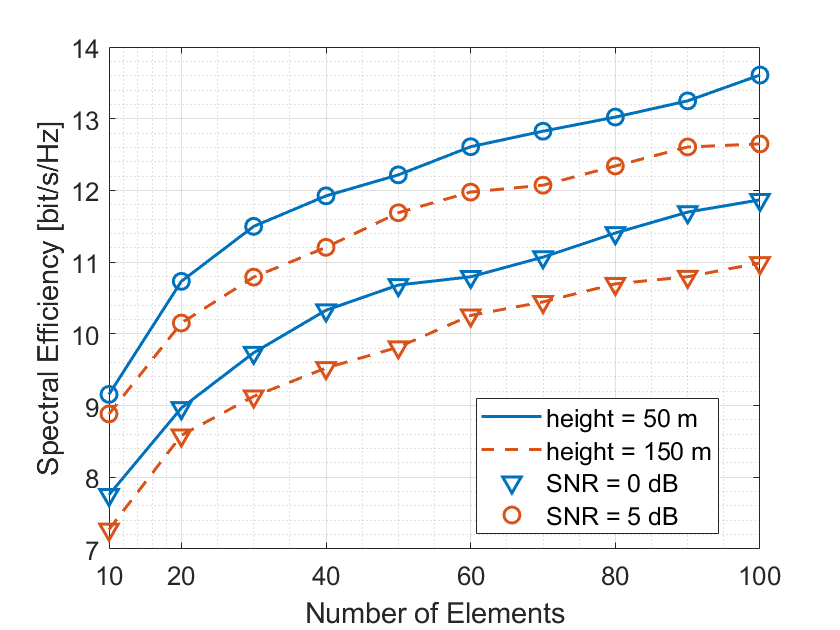}
    \caption{Spectral efficiency vs. number of \ac{ris} elements.}
    \label{fig:se}
\end{figure}

We first consider the spectral efficiency obtained for different numbers of \ac{ris} elements and \ac{uav} flight altitudes. We consider an uplink wireless communication system where the \ac{bs} is equipped with a \ac{ula}. We consider four single-antenna users, placed at the corners of a square centered at the \ac{bs}'s location. We assume that no \ac{los} channel is available from the users to the \ac{bs}, therefore \ac{ris}-enabled \acp{uav} convey the signal to the \ac{bs}. In particular, a single \ac{ris}-enabled \ac{uav} is available for each user. The \ac{bs} uses zero forcing beamforming to mitigate the multi-user interference, and we assume perfect channel knowledge for the beamformer design. While the phase shifts are optimized through non-convex optimization algorithm using coordinate descent, we compute the best location of \acp{uav} via a novel particle swarm optimization algorithm. 
Fig.~\ref{fig:se} shows the average total spectral efficiency versus the number of \ac{ris} elements. Considering two \ac{snr} values (i.e., $0$ and $5$~dB) and two \ac{uav} flying heights (i.e., $50$ and $150$~m). We notice that the \ac{snr} is the main factor that impacts on the system's performance. In fact, while increasing the altitude incurs in a small reduction of the spectral efficiency, increasing the \ac{snr} significantly improves the spectral efficiency. We also notice that an increasing number of elements provides a higher spectral efficiency. This is due to the decrease in the path loss value, as shown in Fig.~\ref{fig:pathloss_all}.

\subsection{Secrecy Application}
Due to the absence of complex cryptographic operations, \ac{pls} techniques are particularly suitable for time-sensitive applications. Thanks to these techniques, communication links can be safeguarded from attacks such as jamming or eavesdropping. We here show the performance of \ac{ris}-enabled \ac{uav}  in a \ac{pls} application. In particular, we assume that an eavesdropper at the \ac{bs}'s side might be able to capture the uplink signals. To assess the secrecy in the overall area, we consider a grid of eavesdropper's locations centered at the \ac{bs} location, and we report the \textit{average secrecy rate}, i.e., the secrecy rate averaged over all grid's locations. This measure is more representative than the classical secrecy rate that considers a specific eavesdropper location. In fact, thanks to the average secrecy rate we are able to characterize the security level with respect to an unknown position of the eavesdropper. 

\begin{figure}[!t]
    \centering
    \includegraphics[width=\columnwidth]{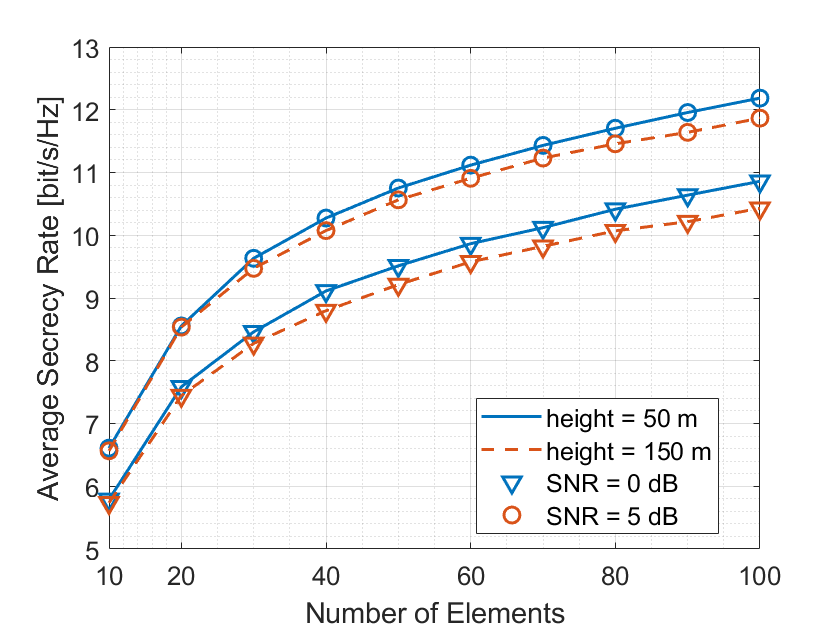}
    \caption{Average secrecy rate vs. number of \ac{ris} elements.  }
    \label{fig:aSec}
\end{figure}

Fig. \ref{fig:aSec} shows the average secrecy rate considering different number of \ac{ris} elements, as a function of the two \ac{snr} and \ac{uav} height values considered for Fig.~\ref{fig:se}. We see that the average secrecy rate increases with the number of reflecting elements. We further notice that, although the considered flight altitudes are very different, they do not affect the network's average secrecy rate. This implies that the path loss is not a determinant factor in the achievable average secrecy rate, as both the channel of the legitimate user and the eavesdropper undergo similar propagation environments.

\section{Future Research Directions}
We believe that the following represent some of the research directions that need to be addressed.
\begin{itemize}
    \item \textit{Design of Suitable Models to Account Physical Constraints}. The shape and size of the \ac{ris} imposes some physical constraints to the \ac{uav} flying mechanism. Researchers shall consider the sail effect that the \ac{ris} may generate in specific meteorological conditions, e.g., high speed wind. These models will be useful for accounting for these effects and compensate them from both a \ac{uav} control and a communication channel point of view.
    \item \textit{Optimization of the Deployment Time of \acp{uav}}. Due to the additional weight imposed by \acp{ris}, we showed that the \ac{uav} flight time is considerably limited. Therefore, practical network deployments of \ac{ris}-enabled \ac{uav} shall consider the timing constraints and orchestrate the available \acp{uav} to minimize the time needed to reach a certain destination and the time in which the \ac{uav} needs to be deployed.
    \item \textit{Machine Learning Controllers}. Recent works proposed the use of machine learning to optimally configure the \ac{ris}'s phase shifts. However, these machine learning algorithm may impose additional constraints on the \ac{uav}'s resources both in terms of memory and power consumption. This will further limit the flight time. Therefore, researchers shall develop low-complexity solutions to guarantee the benefits brought by machine learning algorithms while avoiding the waste of energy resources.
    \item \textit{On the Flight \ac{uav} Charging}. One of the solutions to increase the \ac{uav} flight time is given by charging \acp{uav} while they are deployed. Examples of these solutions include wireless power transfer, where a further \ac{uav} could be deployed to convey the energy required by the deployed \ac{uav} to guarantee the service continuity. Therefore, researchers should develop suitable optimization frameworks accounting for the additional power consumption imposed by the \ac{ris}.
    \item \textit{Design of Suitable Security Measures}. \acp{uav} are vulnerable to multiple types of attacks, both from a network and a physical point of view. Researchers may exploit the advantages brought by \ac{ris}-enabled \acp{uav} to enhance the security of the deployed \acp{uav}. However, such countermeasures shall account for the limitations presented in this paper.
\end{itemize}

\section{Conclusion}
\label{five}
The use of \acp{ris} on \acp{uav} brings several benefits in terms of configurability for a wireless network by increasing the degrees of freedom. However, when considering mounting a \ac{ris} on a \ac{uav} multiple practical factors need to be taken into account. In this paper, we highlighted the most important factors, namely the additional power consumption and weight while carrying an \ac{ris}. We showed how such factors influence the flight time of multiple \ac{uav} models. Furthermore, we discussed the quality of the channel from the transmitter to the receiver through the \ac{ris}-enabled \ac{uav} in terms of path loss, considering the flying altitudes imposed by different countries and showed the path-loss as a function of both carrier frequency and \ac{ris} size. Lastly we showed the performance of this framework in a \ac{pls} scenario, introducing the concept of average secrecy rate. We showed via numerical evaluation that the achievable average secrecy rate increases with the number of \ac{ris} elements. Furthermore, we showed that the flying height and hence the path loss does not heavily influence the achievable average secrecy rate. We conclude that \ac{ris}-enabled \ac{uav} are a feasible solution to enhance the network security. However, the practical considerations we outlined in this paper need to be considered before the deployment of these solutions.

\balance
\bibliographystyle{IEEEtran}

\section*{Biographies}
\begin{IEEEbiographynophoto} {A. Brighente} is a postdoctoral researcher of the Security and Privacy research group (SPRITZ) at the University of Padova.  He received his Ph.D. degree in Information Engineering from the University of Padova in Feb. 2021. From June to November 2019 he was visitor researcher at Nokia Bell Labs, Stuttgart. He published several papers in top-most conferences and journals, and he is also author of a patent. He has been involved in European projects (Ontochain) and company projects (IOTA) with the University of Padova. He is guest editor for IEEE Transactions on Industrial Informatics. His current research interests include security and privacy in cyber-physical systems, vehicular networks, blockchain, and physical layer security.
\end{IEEEbiographynophoto}

\begin{IEEEbiographynophoto}
{M. Conti} is Full Professor at the University of Padua, Italy. He
obtained his Ph.D. from Sapienza University of Rome, Italy, in 2009. He
has been awarded with a Marie Curie Fellowship (2012) and with a
Fellowship by the German DAAD (2013). His research is also funded by
companies, including Cisco, Intel, and Huawei. His main research
interest is in Security and Privacy. He published more than
400 papers in topmost international peer-reviewed journals and
conferences. He is Area Editor-in-Chief for IEEE COMST, and has been
Associate Editor for several journals. He was Program Chair and General
Chair for several international conferences. He is Senior Member of the
IEEE and ACM. He is a member of the Blockchain Expert Panel of the
Italian Government. He is Fellow of the Young Academy of Europe. From
2020, he is Head of Studies of the Master Degree in Cybersecurity at
University of Padua.
\end{IEEEbiographynophoto}

\begin{IEEEbiographynophoto}
{H. Idriss} is a computer and communication engineer. She received her Masters degree from the Lebanese International University in 2018. Her research interests are in 5G technology, vehicular networks, UAV communications, and physical layer security.   
\end{IEEEbiographynophoto}

\begin{IEEEbiographynophoto}
{S. Tomasin} received the Ph.D. degree in Telecommunications Engineering from the University of Padova, Italy, in 2003. In 2002 he joined the University of Padova where he is now Associate Professor. He has been on leave at Philips Research (Eindhoven, Netherlands) in 2002, Qualcomm Research Laboratories (San Diego, California) in 2004, Polytechnic University (Brooklyn, New York) in 2007 and Huawei Mathematical and Algorithmic Sciences Laboratory (Boulogne-Billancourt, France) in 2015. His current research interests include physical layer security and signal processing for wireless communications, with application to 5th generation cellular systems. In 2011-2017 he has been an Editor of the IEEE Transactions of Vehicular Technologies, in 2017-2021 he has been Editor of IEEE Transactions on Signal Processing, and since 2020 he is Editor of the IEEE Transactions on Information Forensics and Security. Since 2011 he is also Editor of EURASIP Journal of Wireless Communications and Networking
\end{IEEEbiographynophoto}

\end{document}